*Research Article*

# On the Nanocommunications at THz Band in Graphene-Enabled Wireless Network-on-Chip

**Quoc-Tuan Vien,[1] Michael Opoku Agyeman,[2] Tuan Anh Le,[1] and Terrence Mak[3]**

[1]*Faculty of Science and Technology, Middlesex University, London NW4 4BT, UK*
[2]*Department of Computing and Immersive Technologies, University of Northampton, Northampton NN2 6JB, UK*
[3]*School of Electronics and Computer Science, University of Southampton, Southampton SO17 1BJ, UK*

Correspondence should be addressed to Quoc-Tuan Vien; q.vien@mdx.ac.uk

One of the main challenges towards the growing computation-intensive applications with scalable bandwidth requirement is the deployment of a dense number of on-chip cores within a chip package. To this end, this paper investigates the Wireless Network-on-Chip (WiNoC), which is enabled by graphene-based nanoantennas (GNAs) in Terahertz frequency band. We first develop a channel model between the GNAs taking into account the practical issues of the propagation medium, such as transmission frequency, operating temperature, ambient pressure, and distance between the GNAs. In the Terahertz band, not only dielectric propagation loss but also molecular absorption attenuation (MAA) caused by various molecules and their isotopologues within the chip package constitutes the signal transmission loss. We further propose an optimal power allocation to achieve the channel capacity. The proposed channel model shows that the MAA significantly degrades the performance at certain frequency ranges compared to the conventional channel model, even when the GNAs are very closely located. More specifically, at transmission frequency of 1 THz, the channel capacity of the proposed model is shown to be much lower than that of the conventional model over the whole range of temperature and ambient pressure of up to 26.8% and 25%, respectively.

## 1. Introduction

Wireless nanocommunication has attracted an extensive investigation of researchers in various fields, such as healthcare, environment, defense, and military services [1, 2]. Advances in nanotechnology with nanomaterials are promising to provide novel solutions for manufacturing various machines in the nanoscale from one to hundred nanometers [3]. However, the data communication between nanomachines over wireless medium via electromagnetic wave is challenging due to their size, complexity, and power consumption [4].

In order to overcome the limitation in wireless nanonetworks, graphene has recently emerged as a new nanomaterial to enable the production of nanoantennas for electromagnetic wave transmission in the THz band [5–7]. Specifically, graphene has been found as a favourable nanomaterial to enable the elaboration of transistors not only providing higher speed but also consuming lower energy compared to the conventional CMOS devices [8–10]. Therefore, graphene has been proposed to build graphene-based plasmonic miniaturized antennas, a.k.a. graphennas [6, 7], to facilitate the communication between nanomachines over wireless medium.

To meet the scalable bandwidth requirement for growing communication- and/or computation-intensive applications, such as emerging multimedia applications, Network-on-Chip (NoC) was initially proposed as an on-chip packet-switched micro-network of wireline routed interconnections [11]. However, the conventional metal based interconnections are insufficient to satisfy the needs of both low latency and high performance as the number of cores increases. Alternative fabrics and architectures, such as photonic NoC [12], nanophotonic NoC [13], three-dimensional (3D) NoC [14], Wireless NoC (WiNoC) [15–19], and hybrid WiNoC [20–23], have been then investigated.

Exploiting the properties of graphene-based nanoantennas (GNAs), WiNoC has adopted the GNAs for employing wireless nanocommunication between cores in the THz band [24–26]. The graphene-enabled WiNoC (GWiNoC) not only helps reduce the propagation delay of intrachip communication but also allows the flexibility and scalability in the chip design due to the inherent characteristics of wireless communications. Furthermore, the THz band can offer enough bandwidth to accommodate massive number of wireless cores in the GWiNoC for emerging System-on-Chip (SoC) design.

The propagation of electromagnetic waves was shown to have a significant effect on the performance of nanocommunications in the THz band [27]. The research challenges in the channel modeling for WiNoC were also discussed in [28]. Therefore, considering the deployment of GNAs in the practical WiNoC, it is crucial to investigate the effects of various propagation environment parameters inside a chip package on the performance of GWiNoC, such as operating temperature, ambient pressure, transmission frequency, and distance between the GNAs. To the best of the authors' knowledge, these issues of the channel modeling in the GWiNoC had not been well investigated.

In this paper, we investigate the communications between GNAs within a GWiNoC environment at THz frequency band. The main contributions of this paper are summarised as follows:

(i) We propose a channel model for GNAs within a GWiNoC environment. At THz frequency band, the total path loss of the signal transmission consists of both dielectric propagation loss (DPL) and molecular absorption attenuation (MAA). Specifically, a within-package reflection channel model is developed taking into account not only line-of-sight and reflected communications but also the transmission medium and built-in material inside a chip. The proposed channel model is regarded as an abstract model for theoretical analysis of various performance metrics allowing us to investigate the impact of the communication environment on the performance of data communication between GNAs.

(ii) The path loss and channel capacity expressions of the proposed model are developed to investigate the impact of the communication environment on the performance of data communication between GNAs.

(iii) We show that the total path loss does not monotonically increase as a function of the frequency but varies over certain frequency ranges, such as 1.21 THz, 1.28 THz, and 1.45 THz, depending on the molecules and their isotopologues within the chip package. Interestingly, the total path loss is shown to decrease over the system electronic noise temperature while it exponentially increases over the ambient pressure applied on the chip. The performance degradation caused by the distance between the GNAs is also shown to be higher than that of the conventional channel model over pure air. (The conventional model used as the baseline is a GWiNoC channel model based on a typical channel model where signals are transmitted over free-space without considering the impact of MAA.)

(iv) We develop an optimal power allocation achieving the derived channel capacity subject to the total power transmission constraint at a GNA. The medium compositions, temperature, and pressure within the chip package are shown to have a significant effect on the total noise temperature and the channel capacity of the GWiNoC.

(v) We show that MAA causes significant degradations in the performance of the nanocommunications within the GWiNoC in comparison with that of the conventional wireless channel model. Specifically, a performance degradation of up to 31.8% is caused by the MAA, even when the GNAs are very closely located of only 0.01 mm. At transmission frequency of 1 THz, the channel capacity of the proposed model is shown to be much lower than that of the conventional model over the whole range of temperature and ambient pressure of up to 26.8% and 25%, respectively. This performance degradation indicates the effectiveness of the proposed channel modeling in capturing the issues of the GNA deployment in the GWiNoC and thus can be regarded as a performance benchmark in the design of GWiNoCs.

The rest of this paper is organised as follows: Section 2 describes the system model of the nanocommunications between two cores in a typical GWiNoC. Section 3 presents the proposed channel model for the GWiNoC. The performance analysis of the proposed channel model is presented in Section 4 where the channel capacity of the GWiNoC is derived and the optimal power allocation is developed to achieve the channel capacity. Numerical results are presented in Section 5 to validate the findings. Finally, Section 6 draws the main conclusions from this paper.

## 2. System Model of Nanocommunications in a GWiNoC

Figure 1 illustrates a typical GWiNoC package where two on-chip cores $\mathscr{C}_\mathcal{T}$ and $\mathscr{C}_\mathcal{R}$ are both equipped with GNAs and are considered as transmitter and receiver cores, respectively. The package is covered by a metal cube box having a longest rectangular side of $d_C$ and a height of $h \ll d_C$. Let $h_T$ and $h_R$ denote the height of the GNAs at $\mathscr{C}_\mathcal{T}$ and $\mathscr{C}_\mathcal{R}$, respectively. The material property of the transmission medium between $\mathscr{C}_\mathcal{T}$ and $\mathscr{C}_\mathcal{R}$ is represented by the relative permittivity (or dielectric constant) $\epsilon_r \geq 1$. (Note that $\epsilon_r = 1$ in vacuum, while $\epsilon_r > 1$ in other materials [29].) A general transmission medium is considered, of which the composition consists of water vapour, carbon dioxide, oxygen, nitrogen, ozone, molecular hydrogen, nitrous oxide, methane, dioxygen, nitrogen oxide, sulfur dioxide, acetylene, ethane, ethylene, methanol, hydrogen cyanide, chloromethane, hydroxyl radical, hydrogen chloride, chlorine monoxide, carbonyl sulfide, formaldehyde, hypochlorous acid, hydrogen peroxide, phosphine, carbonyl fluoride, sulfur hexafluoride, hydrogen sulfide, formic acid, hydroperoxyl radical, chlorine nitrate, nitrosonium ion, hypobromous acid, bromomethane, acetonitrile, carbon tetrafluoride, diacetylene, cyanoacetylene, carbon monosulfide, and sulfur trioxide [30]. These gas compositions are assumed to be time-invariant over the transmission of a data frame and change independently from one frame to another. Let $d$ denote the distance between $\mathscr{C}_\mathcal{T}$ and $\mathscr{C}_\mathcal{R}$. Here, $\mathscr{C}_\mathcal{T}$ and $\mathscr{C}_\mathcal{R}$ are assumed to be located at any positions within the chip package. The detailed investigation of the practical issues of the antenna deployment within the chip can be referred to in [31]. It is noted that the dielectric material causes a higher propagation path loss than the transmission over the pure air. The absorption and resonance of the medium compositions within the chip package should be taken into account in the channel modeling, especially in THz frequency band where various molecules and their isotopologues may cause a considerable MAA [30]. Therefore, the signal transmission between $\mathscr{C}_\mathcal{T}$ and $\mathscr{C}_\mathcal{R}$ in Figure 1 suffers from the path loss caused by not only the DPL but also the MAA.

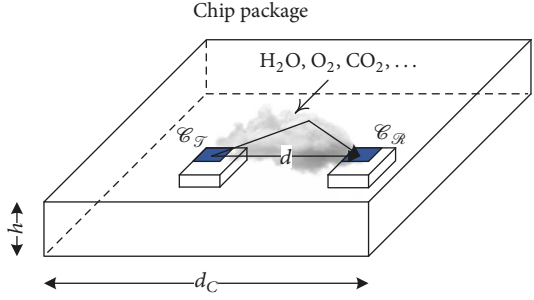

Figure 1: System model of the communications between two cores in a typical GWiNoC package.

Table 1: List of constants.

| Constant name | Symbol | Value |
|---|---|---|
| Avogadro constant | $\zeta_A$ | $6.0221 \times 10^{23}$ mol$^{-1}$ |
| Boltzmann constant | $\zeta_B$ | $1.3806 \times 10^{-23}$ J/K |
| Gas constant | $\zeta_G$ | $8.2051 \times 10^{-5}$ m$^3$atm/K/mol |
| Planck constant | $\zeta_P$ | $6.6262 \times 10^{-34}$ Js |
| Light speed in free-space | $c$ | $2.9979 \times 10^8$ m/s |
| Temperature at standard pressure | $T_p$ | 273.15 K |
| Reference temperature | $T_0$ | 296 K |
| Reference pressure | $p_0$ | 1 atm |

For convenience, the main notations and the well-known constants used in this paper are listed in Notations and Table 1, respectively.

## 3. Proposed Channel Model for GWiNoC

In the proposed channel model for GWiNoC, the total path loss of electromagnetic signal transmitted from $\mathscr{C}_\mathcal{T}$ to $\mathscr{C}_\mathcal{R}$ consists of DPL and MAA. The details are presented as follows.

*3.1. Dielectric Propagation Loss (DPL).* It can be observed in Figure 1 that the data transmission between two cores within the chip package can be carried out via direct line-of-sight (LOS) and/or various reflected, refracted, and scattered waves. In the scope of this work, we consider a specific two-ray propagation model to realise the effects of both the direct and reflection links within the chip package (e.g., in [32]) with the assumption that a single reflection link is dominant over other multipath reflection links. (The considered model is extendible to a general model with multiple rays by adding more rays as in [33].) Therefore, the total received E-field $E_T$ [V/m] at $\mathscr{C}_\mathcal{T}$ consists of the LOS component $E_L$ [V/m] and the reflected component $E_R$ [V/m]. Summing up these two components, we have

$$|E_T| = |E_L + E_R| = 2\frac{E_0 d_0}{d}\sin\left(\frac{\theta}{2}\right), \quad (1)$$

where $E_0$ [V/m] is the E-field at a reference distance $d_0$ [m] and $\theta$ [rad] is the phase difference between the two E-field components. Here, following the same approach as in [32] $\theta$ can be approximated by

$$\theta \approx \frac{4\pi h_T h_R f}{v_p d}, \quad (2)$$

where $h_T$ [m] and $h_R$ [m] denote the height of the GNAs at $\mathscr{C}_\mathcal{T}$ and $\mathscr{C}_\mathcal{R}$, respectively. Here, $v_p$ denotes the phase velocity which is the speed of wave in the dielectric given by [29]

$$v_p = \frac{c}{\sqrt{\epsilon_r}}, \quad (3)$$

where $c = 2.9979 \times 10^8$ m/s is the speed of light in the vacuum.

From (1), (2), and (3), the received power $P_R$ [W] at $\mathscr{C}_\mathcal{R}$ can be computed by

$$P_R = \frac{|E_T|^2 G_R v_p^2}{480\pi^2 f^2}$$
$$= \frac{E_0^2 d_0^2 c^2}{120\pi^2 d^2 f^2 \epsilon_r} G_R \sin^2\left(\frac{2\pi h_T h_R f \sqrt{\epsilon_r}}{cd}\right), \quad (4)$$

where $G_R$ denotes the antenna gain at $\mathscr{C}_\mathcal{R}$. Note that the equivalent isotropically radiated power (EIRP) is given by

$$\text{EIRP} = P_T G_T = \frac{E_0^2 d_0^2 4\pi}{120\pi} = \frac{E_0^2 d_0^2}{30}, \quad (5)$$

where $P_T$ [W] and $G_T$ denote the transmitted power and gain of the GNA at $\mathscr{C}_\mathcal{T}$, respectively. From (4) and (5), $P_R$ can be given by

$$P_R = \frac{P_T G_T G_R}{(2\pi df/c)^2 \epsilon_r} \sin^2\left(\frac{2\pi h_T h_R f \sqrt{\epsilon_r}}{cd}\right). \quad (6)$$

Letting $L_d$ denote the DPL between $\mathscr{C}_{\mathcal{T}}$ and $\mathscr{C}_{\mathcal{R}}$, we then have

$$L_d = \frac{P_T}{P_R} = \left(\frac{2\pi df}{c}\right)^2 \frac{\epsilon_r}{G_T G_R} \csc^2\left(\frac{2\pi h_T h_R f \sqrt{\epsilon_r}}{cd}\right). \quad (7)$$

*Remark 1* (impact of chip material on DPL). It can be noticed from (7) that the DPL (i.e., $L_d$) of the nanocommunications between two on-chip cores depends on not only the physical parameters of the antennas, such as antenna gain (i.e., $G_T$, $G_R$), height (i.e., $h_T$, $h_R$), frequency (i.e., $f$), and distance (i.e., $d$), but also the material of the chip which is reflected by its relative permittivity (i.e., $\epsilon_r$). A further notice is that, when $\epsilon_r \to \infty$, we can deduce $P_R \to 0$ and thus $L_d \to \infty$. This accordingly implies that the chip material has a considerable impact on the performance of the wireless nanocommunications since the data could not be transferred reliably or even totally lost in a high-permittivity medium.

### 3.2. Molecular Absorption Attenuation (MAA).

As the electromagnetic wave at frequency $f$ passes through a transmission medium of distance $d$, there exists a MAA caused by various molecules within the material substance. Let $L_a$ denote the MAA of the data transmission from $\mathscr{C}_{\mathcal{T}}$ to $\mathscr{C}_{\mathcal{R}}$. Applying Beer-Lambert's law to atmospheric measurements, $L_a$ can be determined by

$$L_a = \frac{1}{\tau} = e^{\kappa d}, \quad (8)$$

where $\tau$ is the transmittance of a medium and $\kappa$ [m$^{-1}$] is the medium absorption coefficient. Here, $\kappa$ depends on the composition of the medium (i.e., particular mixture of molecules along the channel) as follows:

$$\kappa = \sum_{i,g} \kappa^{(i,g)}, \quad (9)$$

where $\kappa^{(i,g)}$ [m$^{-1}$] denotes the individual absorption coefficient for the isotopologue $i$ of gas $g$. (For simplicity in representation, the isotopologue $i$ of gas $g$ is hereafter denoted by $(i,g)$.)

Applying radiative transfer theory [34], $\kappa^{(i,g)}$ is given by

$$\kappa^{(i,g)} = \frac{p}{p_0} \frac{T_p}{T_S} Q^{(i,g)} \varsigma^{(i,g)}, \quad (10)$$

where $p$ [atm] is the ambient pressure applied on the chip (the ambient pressure is defined as the pressure of the surrounding medium of the chip which could be varied depending on the application environment of the chip, such as computer chips, sensor chips that could be deployed indoors and/or outdoors, under the water and/or over the air), $T_S$ [K] is the system electronic noise temperature, $p_0 = 1$ atm is the reference pressure, $T_p = 273.15$ K is the temperature at standard pressure, $Q^{(i,g)}$ [mol/m$^3$] is the molecular volumetric density (i.e., number of molecules per volume unit of $(i,g)$), and $\varsigma^{(i,g)}$ [m$^2$/mol] is the absorption cross section of $(i,g)$. Here, $Q^{(i,g)}$ is obtained by the Ideal Gas Law as

$$Q^{(i,g)} = \frac{p}{\zeta_G T_S} q^{(i,g)} \zeta_A, \quad (11)$$

where $\zeta_G = 8.2051 * 10^{-5}$ m$^3$atm/K/mol is the gas constant, $\zeta_A = 6.0221 * 10^{23}$ mol$^{-1}$ is the Avogadro constant, and $q^{(i,g)}$ [%] is the mixing ratio of $(i,g)$.

In (10), $\varsigma^{(i,g)}$ is given by

$$\varsigma^{(i,g)} = S^{(i,g)} \xi^{(i,g)}, \quad (12)$$

where $S^{(i,g)}$ [m$^2$Hz/mol] is the line density for the absorption of $(i,g)$ (i.e., the absorption peak amplitude of $(i,g)$) and $\xi^{(i,g)}$ [Hz$^{-1}$] is spectral line shape of $(i,g)$ determined by

$$\xi^{(i,g)} = \frac{f}{f_c^{(i,g)}} \frac{\tanh\left(\zeta_P v_p f / 2\zeta_B T_S\right)}{\tanh\left(\zeta_P f_c^{(i,g)} / 2\zeta_B T_S\right)} v^{(i,g)}, \quad (13)$$

where $f_c^{(i,g)}$ [Hz] is the resonant frequency of $(i,g)$, $\zeta_P = 6.6262*10^{-34}$ Js is the Planck constant, $\zeta_B = 1.3806*10^{-23}$ J/K is the Boltzmann constant, and $v^{(i,g)}$ [Hz$^{-1}$] is the Van Vleck-Weisskopf asymmetric line shape of $(i,g)$. In (13),

$$f_c^{(i,g)} = f_{c_0}^{(i,g)} + \delta^{(i,g)} \frac{p}{p_0}, \quad (14)$$

where $f_{c_0}^{(i,g)}$ [Hz] is the resonant frequency of $(i,g)$ at reference pressure $p_0 = 1$ atm and $\delta^{(i,g)}$ [Hz] is the linear pressure shift of $(i,g)$. Also, the Van Vleck-Weisskopf asymmetric line shape of $(i,g)$ in (13) can be given by

$$v^{(i,g)} = 100 v_p \frac{\alpha_L^{(i,g)}}{\pi} \frac{f}{f_c^{(i,g)}} \left[ \frac{1}{\left(f - f_c^{(i,g)}\right)^2 + \left(\alpha_L^{(i,g)}\right)^2} \right.$$
$$\left. + \frac{1}{\left(f + f_c^{(i,g)}\right)^2 + \left(\alpha_L^{(i,g)}\right)^2} \right], \quad (15)$$

where $\alpha_L^{(i,g)}$ [Hz] is the Lorentz half width of $(i,g)$. Here, $\alpha_L^{(i,g)}$ is computed by

$$\alpha_L^{(i,g)} = \left[\left(1 - q^{(i,g)}\right)\alpha_0 + q^{(i,g)} \beta^{(i,g)}\right] \frac{p}{p_0} \left(\frac{T_0}{T_S}\right)^\omega, \quad (16)$$

where $\alpha_0$ [Hz] is the broadening coefficient of air, $\beta^{(i,g)}$ [Hz] is the broadening coefficient of $(i,g)$, $T_0 = 296$ K is the reference temperature, and $\omega$ is the temperature broadening coefficient.

### 3.2.1. Proposed Channel Model for GWiNoC.

In summary, let $L$ denote the total path loss for signal transmission from $\mathscr{C}_{\mathcal{T}}$ to $\mathscr{C}_{\mathcal{R}}$. From (7), (8), and (9), the total path loss of the proposed channel model is

$$L = L_d L_a$$
$$= \left(\frac{2\pi df}{c}\right)^2 \frac{\epsilon_r}{G_T G_R} \csc^2\left(\frac{2\pi h_T h_R f \sqrt{\epsilon_r}}{cd}\right) \prod_{i,g} e^{\kappa^{(i,g)} d}, \quad (17)$$

where $\kappa^{(i,g)}$ is determined by (10)–(16).

*Remark 2* (higher path loss due to MAA). In (10), it can be shown that $\kappa^{(i,g)} \geq 0 \; \forall i, g$. From (8) and (17), we have $L_a \geq 1$ and $L \geq L_d$. This accordingly means a higher total path loss is caused by the MAA in the proposed channel model compared to that of the conventional channel model over the pure air with no MAA.

*Remark 3* (environment-aware channel model). The proposed channel model depends on not only the distance between two cores $\mathscr{C}_\mathcal{T}$ and $\mathscr{C}_\mathcal{R}$ but also the absorption of gas molecules, the system electronic temperature, and the ambient pressure applied on the chip. In fact, from (10)–(16), the individual absorption coefficient of the isotopologue $i$ of gas $g$ (i.e., $\kappa^{(i,g)}$) is shown to be dependent but not monotonically varied over $f$. It can also be shown that the total path loss (i.e., $L$) in (17) monotonically decreases over the system temperature (i.e., $T_S$) but exponentially increases over the ambient pressure (i.e., $p$).

From Remarks 1, 2, and 3 in deriving the path loss, the communication environment is shown to have a considerable impact on the performance of the nanocommunications between GNAs within a GWiNoC at THz frequency band. This accordingly reflects the novelty of our proposed channel model which not only allows us to analyse the performance of the GWiNoC which will be shown in the following section, but also implies the design requirements of GWiNoC as well as its feasibility and applicability when employing the GNAs within a chip at THz band.

### 3.3. Link Budget Analysis.
Taking into account all the antenna gains and losses caused by both DPL and MAA, the received power at core $\mathscr{C}_\mathcal{R}$, in dB, can be derived from (17) as

$$
\begin{aligned}
P_R \,[\text{dBW}] &= P_T \,[\text{dBW}] + G_T \,[\text{dB}] + G_R \,[\text{dB}] \\
&\quad - 10 \log_{10} \epsilon_r \\
&\quad - 20 \log_{10} \left[ \frac{2\pi d f}{c} \csc\left( \frac{2\pi h_T h_R f \sqrt{\epsilon_r}}{cd} \right) \right] \\
&\quad - 10 \log_{10} \left( \prod_{i,g} e^{\kappa^{(i,g)} d} \right) \\
&\approx P_T \,[\text{dBW}] + G_T \,[\text{dB}] + G_R \,[\text{dB}] \\
&\quad - 10 \log_{10} \epsilon_r \\
&\quad - 20 \log_{10} \left[ \frac{2\pi d f}{c} \csc\left( \frac{2\pi h_T h_R f \sqrt{\epsilon_r}}{cd} \right) \right] \\
&\quad - 4.343 \left( \sum_{i,g} \kappa^{(i,g)} d \right).
\end{aligned} \quad (18)
$$

Notice that the power received at $\mathscr{C}_\mathcal{R}$ in (18) reflects on the received signal-to-noise ratio (SNR), which is an important measure for evaluating the effectiveness of a communication system as will be shown in the following section.

## 4. Channel Capacity of GWiNoC

In this section, the channel capacity of GWiNoC is derived for the proposed channel model. (Given the link budget analysis in Section 3.3, the bit error rate (BER) can be derived depending on modulation schemes. Although the BER analysis is worth a detailed evaluation, it is beyond the scope of this work in which we aim to evaluate the maximum achievable rate at which the data can be reliably transmitted over the wireless communication channel between two cores at THz band.) We have the following findings.

**Theorem 4.** *The channel capacity, in bits/s, of a nanocommunication system between two THz GNAs within the chip package is given by*

$$
C = \max_{\sum_{k=1}^{K} P_k \leq P_T} \sum_{k=1}^{K} \Delta f \cdot \log_2 \left[ 1 + \frac{P_k G_T G_R \sin^2\left( 2\pi h_T h_R f_k \sqrt{\epsilon_r}/cd \right)}{\zeta_B \epsilon_r \left( 2\pi d f_k/c \right)^2 \Delta f \left[ (T_S + T_0) \prod_{i,g} e^{\kappa_k^{(i,g)} d} - T_0 \right]} \right], \quad (19)
$$

*where $K$ is the number of subbands in the total channel bandwidth of $B$ [Hz], $\Delta f = B/K$ [Hz] is the width of each subband, $P_k$ [W] is the power allocated for the kth subband, $f_k$ [Hz] is the centre frequency of the kth subband, and $\kappa_k^{(i,g)}$ is the individual absorption coefficient for the isotopologue ith of gas gth at frequency $f_k$.*

*Proof.* See Appendix A. □

As expressed in Theorem 4, the channel capacity of the GWiNoC is a function of the power allocated at various subbands (i.e., $P_1, P_2, \ldots, P_K$). Therefore, it is crucial to solve the optimisation problem in (19) subject to the power constraint at $\mathscr{C}_\mathcal{T}$ (i.e., $\sum_{k=1}^{K} P_k \leq P_T$). We have the following proposition.

**Proposition 5.** *The channel capacity of the GWiNoC can be obtained as*

$$
C = \sum_{k=1}^{K} \Delta f \log_2 \left[ 1 + \frac{(\vartheta - \Psi_k)^+}{\Psi_k} \right], \quad (20)
$$

*where $(x)^+ = \max\{0, x\}$,*

$$
\Psi_k = \frac{\zeta_B \epsilon_r}{G_T G_R} \left( \frac{2\pi d f_k}{c} \right)^2 \Delta f \csc^2\left( \frac{2\pi h_T h_R f_k \sqrt{\epsilon_r}}{cd} \right) \cdot \left[ (T_S + T_0) \prod_{i,g} e^{\kappa_k^{(i,g)} d} - T_0 \right], \quad (21)
$$

*and $\vartheta$ can be found by water-filling method such that*

$$
\sum_{k=1}^{K} (\vartheta - \Psi_k)^+ = P_T. \quad (22)
$$

*Proof.* See Appendix B. □

Considering the scenario of employing GNAs of a very small size compared to the distance between two on-chip cores, the channel capacity of the GWiNoC can be approximately computed as in the following corollary.

**Corollary 6.** *When $h_T \ll d$, $h_R \ll d$, $d \to 0$, and $G_T = G_R = 1$, the channel capacity of the GWiNoC can be determined by*

$$C \approx \sum_{k=1}^{K} \Delta f \log_2 \left[ 1 + \frac{(\varphi - \Phi_k)^+}{\Phi_k} \right], \quad (23)$$

*where*

$$\Phi_k = \frac{\zeta_B d^4 \Delta f \left[ T_S + (T_S + T_0) \kappa_k d \right]}{h_T^2 h_R^2} \quad (24)$$

*and $\varphi$ is chosen such that*

$$\sum_{k=1}^{K} (\varphi - \Phi_k)^+ = P_T. \quad (25)$$

*Proof.* As $h_T \ll d$, $h_R \ll d$, and $d \to 0$, applying Maclaurin series [35, eq. (0.318.2)], it can be approximated that

$$\sin^2 \left( \frac{2\pi h_T h_R f_k \sqrt{\epsilon_r}}{cd} \right) \approx \left( \frac{2\pi h_T h_R f_k \sqrt{\epsilon_r}}{cd} \right)^2,$$

$$\prod_{i,g} e^{\kappa_k^{(i,g)} d} \approx 1 + \sum_{i,g} \kappa_k^{(i,g)} d = 1 + \kappa_k d. \quad (26)$$

From (21), after some mathematical manipulations, we have

$$\Psi_k \approx \Phi_k = \frac{\zeta_B d^4 \Delta f \left[ T_S + (T_S + T_0) \kappa_k d \right]}{h_T^2 h_R^2}. \quad (27)$$

Substituting (27) into (20) with the assumption of $G_T = G_R = 1$, the corollary is proved. □

*Remark 7* (impact of molecules within the material substance of a chip). It can be noticed in Proposition 5 and Corollary 6 that the communications environment has a significant impact on the channel capacity of GWiNoC operating at THz band. In particular, besides the physical parameters of antennas, operating frequency, temperature, and ambient pressure, as shown in (23) and (24), the medium absorption coefficient (i.e., $\kappa_k$), which is dependent on the mixture of molecules and their composition inside a chip, also affects the performance of the nanocommunications between the GNAs.

## 5. Numerical Results

In this section, the performance evaluation of the proposed channel model for the nanocommunications in GWiNoC in THz band is carried out through numerical results and compared against the conventional channel model. In the conventional channel model, only pure air with no MAA (e.g., two-ray channel model in [32]) is considered. In our model (see Figure 1), the medium consists of water vapour and various gas compositions, like carbon dioxide, oxygen, nitrogen, and so forth, parameters of which are obtained from the HITRAN database [30]. Additionally, the power for subbands is allocated as in Proposition 5. The impacts of the transmission medium and various channel environment parameters on the performance of GWiNoC in terms of path loss and channel capacity are evaluated to realise the effectiveness of the proposed channel model over the conventional approach. (In the conventional channel model, only DPL is taken into account, and thus it can be regarded as a special case of our channel model where we consider both the DPL and MAA.)

*5.1. Impacts of High Transmission Frequency.* Let us first investigate the impacts of transmission frequency of GNAs on the channel modeling for GWiNoC. Figures 2 and 3 sequentially plot the total path loss and channel capacity of two considered channel models versus the transmission frequency at $\mathscr{C}_\mathscr{T}$. Two cores $\mathscr{C}_\mathscr{T}$ and $\mathscr{C}_\mathscr{R}$ are located within the chip package having the longest side (i.e., $d_C$) of 20 mm and the height (i.e., $h$) of 1 mm. In Figure 2, the distance between $\mathscr{C}_\mathscr{T}$ and $\mathscr{C}_\mathscr{R}$ (i.e., $d$) is set as $d = \{0.1, 1, 10, 20\}$ mm. Each core deploys a GNA having a height of 0.02 mm (i.e., $h_T = h_R = 0.02$ mm). The transmission frequency of the GNA (i.e., $f$) is assumed to vary in the range from 1 to 3 THz. The system electronic noise temperature (i.e., $T_S$) is 296 K and the ambient pressure applied on the chip (i.e., $p$) is 1 atm. It can be observed in Figure 2 that the proposed channel model results in a higher total path loss compared to the conventional channel model and a longer distance between the cores causes a considerably increased path loss. Also, the total path loss is not shown to monotonically increase at the THz frequency band due to the fact that the MAA is caused by isotopologues of gases having various absorption coefficients at various frequencies. For example, the MAA causes a considerably higher path loss at 1.21 THz, 1.28 THz, 1.45 THz, and so forth. In fact, such increased path loss is caused by the molecular absorption of isotopologues of various gases since they have different absorption coefficients at different frequencies (detailed molecular spectroscopic data of different molecules can be referred to in [30]).

Although the total path loss in our proposed model is not much higher than that in the conventional model, it has a significant impact on the channel capacity at the THz band. This indeed can be seen in Figure 3 where the channel capacity achieved in the proposed model is much lower than that of the conventional model. Specifically, at least $0.0005 \times 10^{12}$ bits/s (i.e., 500 Megabits/s) are lost due to the MAA throughout the frequency range from 1 THz to 2 THz. These observations confirm the statements in Remarks 2 and 3 regarding the effectiveness of the proposed channel model with environment-aware property. The proposed channel model can be therefore used to represent the practical scenario of GWiNoC implementation at the THz band where some frequencies (e.g., 1.21 THz, 1.28 THz, and 1.45 THz) causing a significant capacity loss should be avoided in the GNA design.

*5.2. Impacts of Operating Temperature.* The impacts of operating temperature of the chip on the channel modeling of the GWiNoC are shown in Figures 4 and 5, where the total path loss and channel capacity of the proposed and the conventional channel models are plotted against the system electronic noise temperature (i.e., $T_S$) with respect to different values of frequency band (i.e., $f = 1$ THz, $f = 1.2$ THz, and $f = 1.5$ THz). The size of the chip package, the distance between $\mathcal{C}_\mathcal{T}$ and $\mathcal{C}_\mathcal{R}$, the height of the GNAs, and the ambient pressure are similarly set as those in Figures 2 and 3. It can be observed that the system temperature does not have any effects in the path loss of the conventional channel model, while the total path loss in the proposed channel model is shown to decrease as the temperature increases at all frequency bands. This observation accordingly verifies the statement in Remark 3 on the monotonically decreasing total path loss over the system temperature due to the MAA.

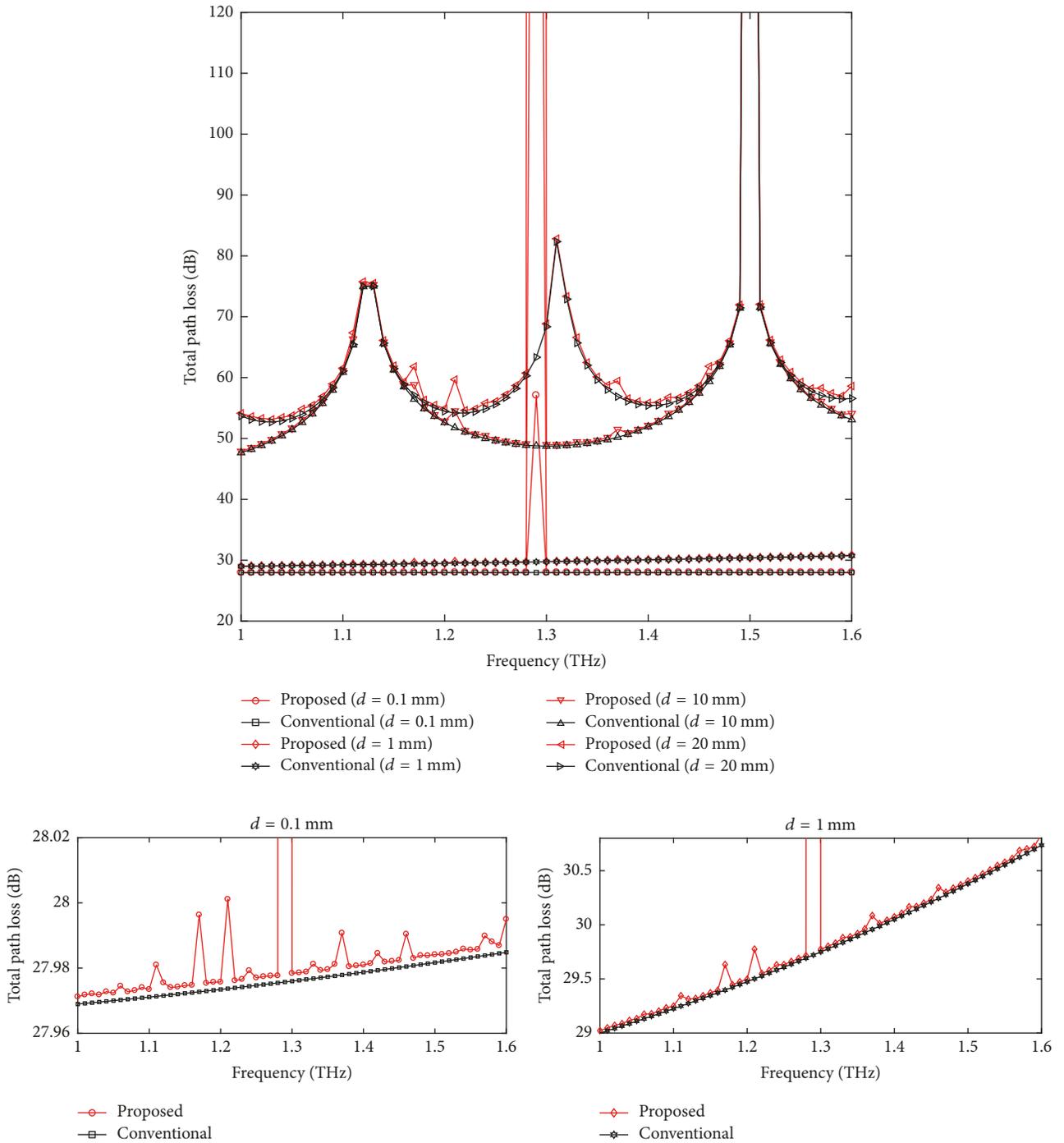

Figure 2: Total path loss versus frequency.

Regarding the performance of the GWiNoC, it can be observed in Figure 5 that the channel capacity of the proposed model is much lower than that of the conventional model at all temperature range. Specifically, 400 Gigabits/s and 1.1 Tetrabits/s are reduced at frequencies 1.2 THz and 1 THz, respectively, in the proposed model when operating at temperature $T_S = 300$ K. This again reflects the effects of various gas compositions in the medium causing a considerably reduced channel capacity of up to 26.8% in the nanocommunications within a chip.

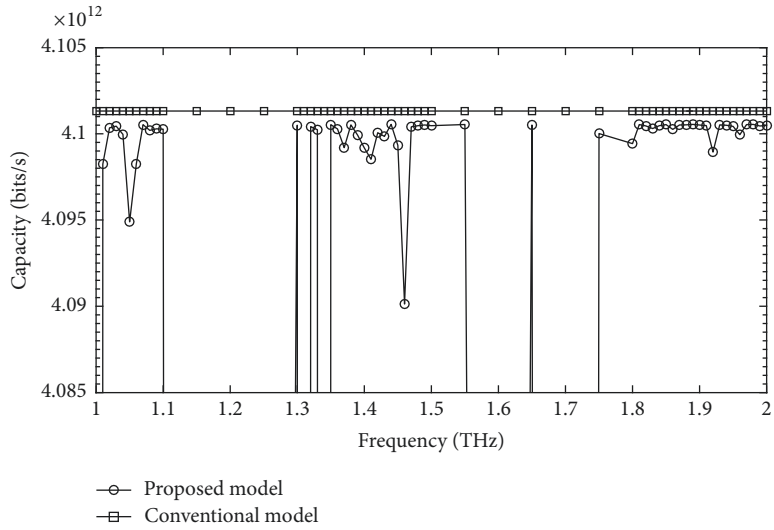

Figure 3: Channel capacity versus transmission frequency.

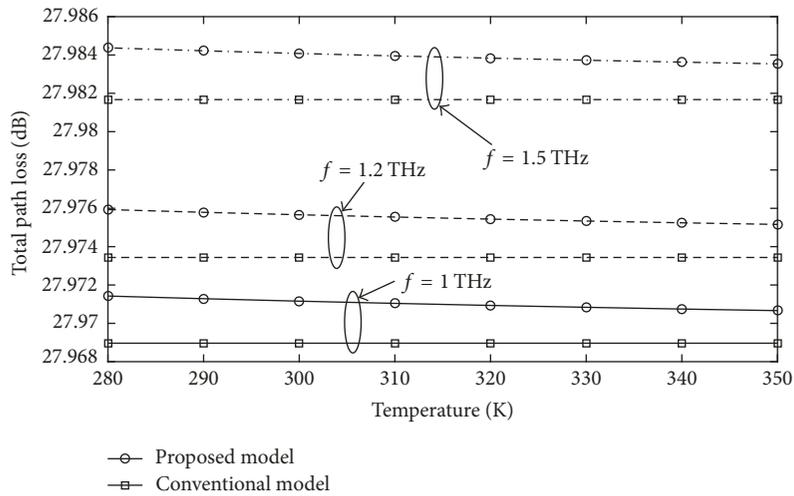

Figure 4: Total path loss versus system electronic noise temperature at different frequencies.

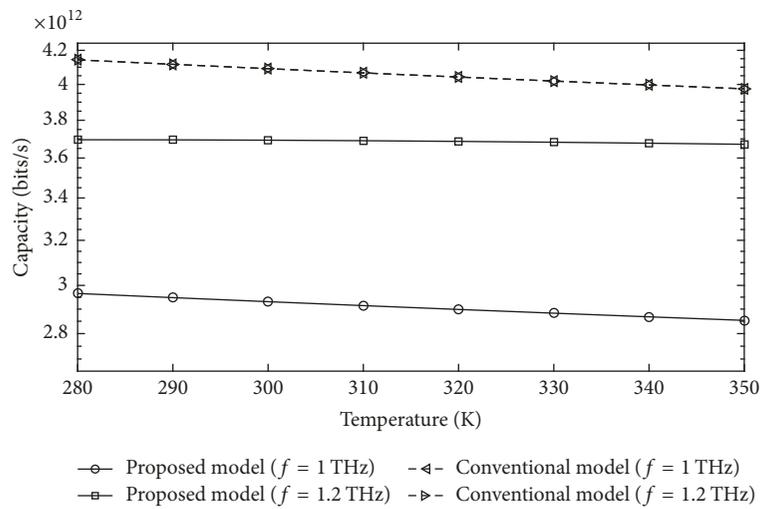

Figure 5: Channel capacity versus system electronic noise temperature.

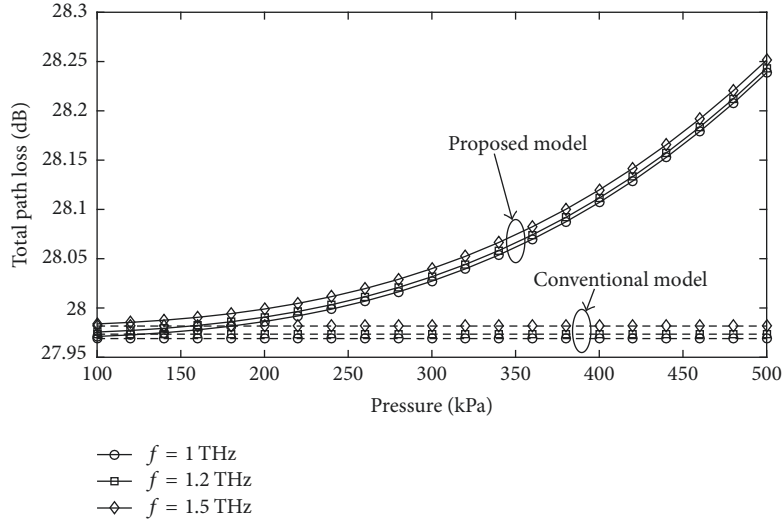

Figure 6: Total path loss versus ambient pressure at different frequencies.

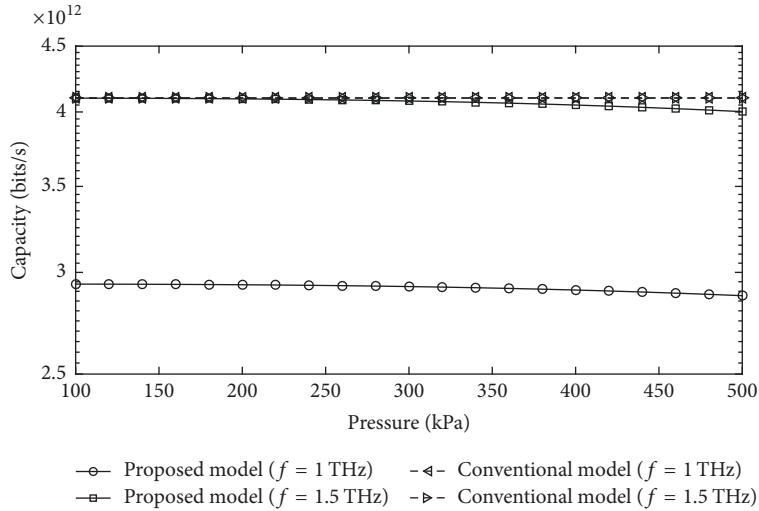

Figure 7: Channel capacity versus ambient pressure.

*5.3. Impacts of Ambient Pressure.* Taking into consideration ambient pressure in GWiNoC, Figures 6 and 7 plot the total path loss and channel capacity of various channel models versus the ambient pressure (i.e., $p$ in kPa) applied on the chip package. (Note that 1 atm = 101.325 kPa.) The GNAs are assumed to operate at frequency $f = \{1, 1.2, 1.5\}$ THz. Similar to Figure 2, the size of the chip package, the distance between $\mathscr{C}_{\mathcal{T}}$ and $\mathscr{C}_{\mathcal{R}}$, the height of the GNAs, and the system electronic noise temperature are set as $d_C = 20$ mm, $h = 1$ mm, $d = 0.1$ mm, $h_T = h_R = 0.02$ mm, and $T_S = 296$ K. It can be seen in Figure 6 that the total path loss in the conventional channel model is independent of the ambient pressure. However, the total path loss in the proposed channel model for practical GWiNoC is shown to exponentially increase as the ambient pressure increases, which confirms the claim of the exponentially increased total path loss over the ambient pressure in Remark 3.

Furthermore, in Figure 7, the capacity of the proposed channel model is shown to be lower than that of the conventional model over the whole range of pressure, especially a degraded performance of up to 25%, when $f = 1$ THz. This achieved performance is also consistent with the achieved performance in Figures 3 and 5.

*5.4. Impacts of Transmission Distance.* Considering the impacts of distance between two cores on the performance of GWiNoC, in Figure 8, the channel capacity of various channel models is plotted as a function of the transmission distance between two cores $\mathscr{C}_\mathscr{T}$ and $\mathscr{C}_\mathscr{R}$ (i.e., $d$). Both flat and water-filling based approaches are considered for power allocation at subbands. The GNAs are assumed to operate at frequency $f = 1$ THz and the distance $d$ is assumed to vary in the range $[10 : 100]$ $\mu$m, while the other simulation parameters are similarly set as in Figure 3. It can be observed that the channel capacity in both the proposed and the conventional channel models decreases as the distance increases, which can be verified from (19) in Theorem 4. Also, in the proposed model, the water-filling based power allocation is shown to provide an improved performance of up to 200 Gigabits/s compared to the flat power allocation, while there is not much difference of the channel capacity of the conventional model using these techniques. Furthermore, the proposed model is shown to achieve a much lower channel capacity compared to the conventional model. For instance, the channel capacity of the proposed model is of up to 31.8% lower than that of the conventional model even when the distance between two cores is only 0.01 mm. This is indeed caused by the introduction of the MAA in the proposed model.

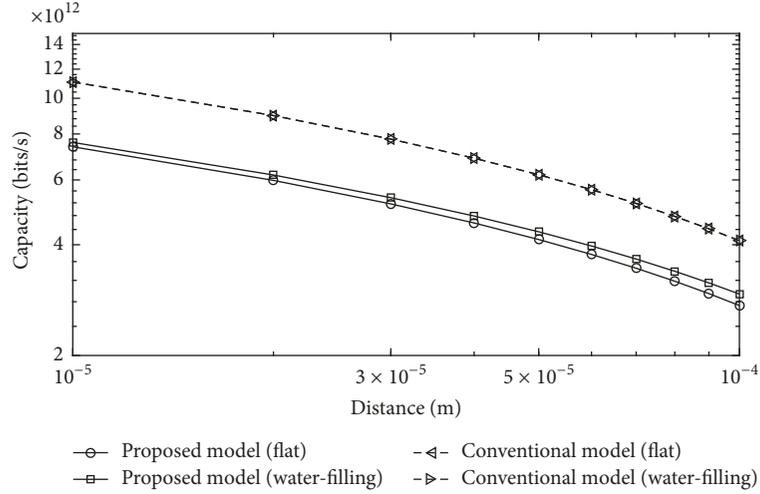

Figure 8: Channel capacity versus distance between two GNAs.

## 6. Conclusions

In this paper, we have proposed an efficient channel model for nanocommunications via GNAs in GWiNoC taking into account the practical issues of the propagation medium within the chip package. It has been shown that MAA has a considerable effect on the performance of the practical GWiNoC, especially in THz frequency band. Specifically, the MAA has been shown to cause a very high path loss at certain frequencies (e.g., 1.21 THz, 1.28 THz, and 1.45 THz) rather than monotonically increasing over the whole frequency range as in the conventional pure-air channel model with only DPL. Additionally, the total path loss has been shown to decrease as the system electronic noise temperature increases, while it exponentially increases as the ambient pressure applied on the chip increases. As in the conventional channel model, similar impacts of the distance between two cores on the performance have been verified in the proposed channel model. Furthermore, it has also been shown that the proposed channel model results in a lower channel capacity compared to the conventional channel model, which reflects the practical issues of the nanocommunications in THz band in GWiNoC. Specifically, at least 500 Megabits/s is lower throughout the frequency range from 1 THz to 2 THz, up to 26.8% and 25% of the channel capacity are reduced over the whole range of temperature and ambient pressure, respectively, and also 3.5 Terabits/s of up to 31.8% is reduced when the distance between two GNAs is 0.01 mm. For future work, we will investigate a general propagation model with multiple rays. In addition, we will examine the practical issues when deploying the GNAs with different layouts and cross section of chip package.

## Appendix

## A. Proof of Theorem 4

As the signal-to-noise ratio (SNR) is required for evaluating the achievable capacity of a communication system, we first derive the total noise power of the nanocommunication between two GNAs. At a frequency $f$ [Hz], the total noise temperature at $\mathscr{C}_\mathscr{R}$ located at $d$ [m] from $\mathscr{C}_\mathscr{T}$ (i.e., $T_{\text{tot}}$ [K]) consists of system electronic noise temperature (i.e., $T_S$ [K]), molecular absorption noise temperature (i.e., $T_M$ [K]), and other noise source temperature (i.e., $T'$ [K]); that is,

$$T_{\text{tot}} = T_S + T_M + T'. \tag{A.1}$$

Assuming that $T_S + T_M \gg T'$ $\forall f, d$, we have

$$T_{\text{tot}} \approx T_S + T_M. \tag{A.2}$$

Here, $T_M$ is caused by the molecules within the transmission medium and thus can be expressed via the transmittance of the medium as

$$T_M = T_0(1 - \tau) = T_0\left(1 - \prod_{i,g} e^{-\kappa^{(i,g)}d}\right). \tag{A.3}$$

Substituting (A.3) into (A.2), we obtain

$$T_{\text{tot}} \approx T_S + T_0 \left(1 - \prod_{i,g} e^{-\kappa^{(i,g)} d}\right). \quad (A.4)$$

The total noise power at $\mathscr{C}_\mathscr{R}$ given transmission bandwidth $B$ is therefore given by

$$P_N(d) = \zeta_B \int_B T_{\text{tot}} df. \quad (A.5)$$

Note that the THz channel is highly frequency-selective and the molecular absorption noise is nonwhite. Therefore, we can divide the total bandwidth $B$ into $K$ narrow subbands and regard them as parallel channels. Subject to the transmission power constraint at $\mathscr{C}_\mathscr{T}$ (i.e., $\sum_{k=1}^{K} P_k \leq P_T$), the channel capacity, in bits/s, of the nanocommunications between the GNAs can be expressed by

$$C = \max_{\sum_{k=1}^{K} P_k \leq P_T} \sum_{k=1}^{K} \Delta f \log_2 \left[1 + \frac{P_k}{\zeta_B L_k T_{\text{tot},k} \Delta f}\right], \quad (A.6)$$

where $\Delta f = B/K$ [Hz] is the width of subband, $f_k$ [Hz] is the centre frequency of the $k$th subband, $P_k$ is the power allocated for the $k$th subband, $L_k$ is the total path loss, and $T_{\text{tot},k}$ is the total noise temperature at $f_k$. Substituting (17) and (A.4) into (A.6), we obtain (19). The theorem is proved.

## B. Proof of Proposition 5

Let $\lambda$ be the Lagrange multiplier associated with the power constraint $\sum_{k=1}^{K} P_k - P_T \leq 0$; the Lagrangian of (19) can be formed as

$$\mathscr{L}(\lambda, P_k) = \sum_{k=1}^{K} \Delta f \log_2 \left[1 + \frac{P_k G_T G_R \sin^2\left(2\pi h_T h_R f_k \sqrt{\epsilon_r}/cd\right)}{\zeta_B \epsilon_r (2\pi d f_k/c)^2 \Delta f \left[(T_S + T_0) \prod_{i,g} e^{\kappa_k^{(i,g)} d} - T_0\right]}\right] \\ + \lambda \left(\sum_{k=1}^{K} P_k - P_T\right). \quad (B.1)$$

By denoting $\Psi_k$ as in (21), we can rewrite (B.1) as

$$\mathscr{L}(\lambda, P_k) = \sum_{k=1}^{K} \Delta f \log_2 \left(1 + \frac{P_k}{\Psi_k}\right) \\ + \lambda \left(\sum_{k=1}^{K} P_k - P_T\right). \quad (B.2)$$

Differentiating $\mathscr{L}(\lambda, P_k)$ with respect to $P_k$, $k = 1, 2, \ldots, K$, we have

$$\frac{\partial \mathscr{L}(\lambda, P_k)}{\partial P_k} = \frac{\Delta f}{\ln 2} \frac{1}{P_k + \Psi_k} + \lambda. \quad (B.3)$$

Solving $\partial \mathscr{L}(\lambda, P_k)/\partial P_k = 0$, we obtain

$$P_k = -\frac{\Delta f}{\lambda \ln 2} - \Psi_k = \vartheta - \Psi_k, \quad (B.4)$$

where $\vartheta \triangleq -\Delta f/(\lambda \ln 2)$.

Since $P_k \geq 0$, $k = 1, 2, \ldots, K$, the solution in (B.4) is

$$P_k = (\vartheta - \Psi_k)^+, \quad (B.5)$$

where $(x)^+ \triangleq \max\{0, x\}$ and $\vartheta$ can be solved by utilising the power constraint at $\mathscr{C}_\mathscr{T}$ as in water-filling approach. That is,

$$\sum_{k=1}^{K} (\vartheta - \Psi_k)^+ = P_T. \quad (B.6)$$

This completes the proof.

## Notations

| | |
|---|---|
| $d$ [m]: | Distance between two GNAs |
| $h$ [m]: | Height of the chip package |
| $h_T, h_R$ [m]: | Height of the GNAs at $\mathscr{C}_\mathscr{T}$ and $\mathscr{C}_\mathscr{R}$ |
| $f, B$ [Hz]: | Transmission frequency and channel bandwidth |
| $p$ [atm]: | Ambient pressure applied on chip |
| $T_S, T_M$, and $T'$ [K]: | System electronic, molecular absorption, and other noise source temperature, respectively |
| $L_d, L_a$, and $L$: | DPL, MAA, and total path loss, respectively |
| $P_T$ and $P_R$ [W]: | Transmitted power and received power |
| $G_T$ and $G_R$: | Transmitter antenna gain and receiver antenna gain |
| $v_p$ [m/s]: | Phase velocity |
| $\epsilon_r$: | Relative permittivity of material |
| $\tau$: | Transmittance of medium |
| $\kappa$: | Medium absorption coefficient |
| $(i, g)$: | Isotopologue $i$ of gas $g$ |
| $\kappa^{(i,g)}$: | Individual absorption coefficient of $(i, g)$ |
| $Q^{(i,g)}$ [mol/m$^3$]: | Molecular volumetric density of $(i, g)$ |
| $\varsigma^{(i,g)}$ [m$^2$/mol]: | Absorption cross section of $(i, g)$ |
| $q^{(i,g)}$ [%]: | Mixing ratio of $(i, g)$ |
| $S^{(i,g)}$ [m$^2$Hz/mol]: | Line density for the absorption of $(i, g)$ |
| $\xi^{(i,g)}$ [Hz$^{-1}$]: | Spectral line shape of $(i, g)$ |
| $f_c^{(i,g)}, f_{c_0}^{(i,g)}$ [Hz]: | Resonant frequency of $(i, g)$ and resonant frequency at $p_0 = 1$ atm |
| $\nu^{(i,g)}$ [Hz$^{-1}$]: | Van Vleck-Weisskopf asymmetric line shape [36] |
| $\delta^{(i,g)}$ [Hz]: | Linear pressure shift of $(i, g)$ |
| $\alpha_L^{(i,g)}$ [Hz]: | Lorentz half width of $(i, g)$ [36] |
| $\alpha_0, \beta^{(i,g)}$ [Hz]: | Broadening coefficient of air and $(i, g)$ respectively |
| $\omega$: | Temperature broadening coefficient. |

## Conflicts of Interest

The authors declare that there are no conflicts of interest regarding the publication of this paper.

## References

[1] P. Wang, J. M. Jornet, M. G. Abbas Malik, N. Akkari, and I. F. Akyildiz, "Energy and spectrum-aware MAC protocol for perpetual wireless nanosensor networks in the Terahertz Band," *Ad Hoc Networks*, vol. 11, no. 8, pp. 2541–2555, 2013.

[2] I. F. Akyildiz and J. M. Jornet, "Electromagnetic wireless nanosensor networks," *Nano Communication Networks*, vol. 1, no. 1, pp. 3–19, 2010.

[3] I. F. Akyildiz and J. M. Jornet, "The Internet of nano-things," *IEEE Wireless Communications*, vol. 17, no. 6, pp. 58–63, 2010.

[4] S. Balasubramaniam and J. Kangasharju, "Realizing the internet of nano things: challenges, solutions, and applications," *Computer*, vol. 46, no. 2, pp. 62–68, 2013.

[5] I. Llatser, A. Cabellos-Aparicio, E. Alarcon et al., "Scalability of the channel capacity in graphene-enabled wireless communications to the nanoscale," *IEEE Transactions on Communications*, vol. 63, no. 1, pp. 324–333, 2015.

[6] J. M. Jornet and I. F. Akyildiz, "Graphene-based plasmonic nano-antenna for terahertz band communication in nanonetworks," *IEEE Journal on Selected Areas in Communications*, vol. 31, no. 12, pp. 685–694, 2013.

[7] J. M. Jornet and I. F. Akyildiz, "Graphene-based nano-antennas for electromagnetic nanocommunications in the terahertz band," in *Proceedings of the EuCAP 2010*, pp. 1–5, Barcelona, Spain, April 2010.

[8] Y. Wu, K. A. Jenkins, A. Valdes-Garcia et al., "State-of-the-art graphene high-frequency electronics," *Nano Letters*, vol. 12, no. 6, pp. 3062–3067, 2012.

[9] A. K. Geim, "Graphene: status and prospects," *Science*, vol. 324, no. 5934, pp. 1530–1534, 2009.

[10] T. Otsuji, S. Tombet, A. Satou, M. Ryzhii, and V. Ryzhii, "Terahertz-wave generation using graphene: Toward new types of terahertz lasers," *IEEE Journal of Selected Topics in Quantum Electronics*, vol. 19, no. 1, p. 8400209, 2013.

[11] T. Bjerregaard and S. Mahadevan, "A survey of research and practices of network-on-chip," *ACM Computing Surveys*, vol. 38, no. 1, 2006.

[12] A. Shacham, K. Bergman, and L. P. Carloni, "Photonic networks-on-chip for future generations of chip multiprocessors," *Institute of Electrical and Electronics Engineers. Transactions on Computers*, vol. 57, no. 9, pp. 1246–1260, 2008.

[13] Y. Xu and S. Pasricha, "Silicon nanophotonics for future multicore architectures: opportunities and challenges," *IEEE Design & Test*, vol. 31, no. 5, pp. 9–17, 2014.

[14] M. O. Agyeman, A. Ahmadinia, and A. Shahrabi, "Heterogeneous 3d network-on-chip architectures: Area and power aware design techniques," *Journal of Circuits, Systems and Computers*, vol. 22, no. 4, Article ID 1350016, 2013.

[15] S. Deb, A. Ganguly, P. P. Pande, B. Belzer, and D. Heo, "Wireless NoC as interconnection backbone for multicore chips: Promises and challenges," *IEEE Journal on Emerging and Selected Topics in Circuits and Systems*, vol. 2, no. 2, pp. 228–239, 2012.

[16] D. W. Matolak, A. Kodi, S. Kaya, D. Ditomaso, S. Laha, and W. Rayess, "Wireless networks-on-chips: Architecture, wireless channel, and devices," *IEEE Wireless Communications*, vol. 19, no. 5, pp. 58–65, 2012.

[17] S. Abadal, M. Iannazzo, M. Nemirovsky, A. Cabellos-Aparicio, H. Lee, and E. Alarcón, "On the area and energy scalability of wireless network-on-chip: A model-based benchmarked design space exploration," *IEEE/ACM Transactions on Networking*, vol. 23, no. 5, pp. 1501–1513, 2015.

[18] M. O. Agyeman, Q.-T. Vien, and T. Mak, "An analytical channel model for emerging wireless networks-on-chip," in *Proceedings of the IEEE/IFIP EUC 2016*, Paris, France, August 2016.

[19] M. O. Agyeman, Q. T. Vien, G. Hill, S. Turner, and T. Mak, "An efficient channel model for evaluating wireless NoC architectures," in *Proceedings of the SBAC-PADW 2016*, pp. 85–90, October 2016.

[20] P. Pande, A. Ganguly, K. Chang, and C. Teuscher, "Hybrid wireless network on chip: a new paradigm in multi-core design," in *Proceedings of the NoCArc'09 Workshop*, pp. 71–76, New York, NY, USA, December 2009.

[21] A. Ganguly, K. Chang, S. Deb, P. P. Pande, B. Belzer, and C. Teuscher, "Scalable hybrid wireless network-on-chip architectures for multicore systems," *Institute of Electrical and Electronics Engineers. Transactions on Computers*, vol. 60, no. 10, pp. 1485–1502, 2011.

[22] M. O. Agyeman, J.-X. Wan, Q.-T. Vien et al., "On the Design of Reliable Hybrid Wired-Wireless Network-on-Chip Architectures," in *Proceedings of the 9th IEEE International Symposium on Embedded Multicore/Manycore SoCs, MCSoC 2015*, pp. 251–258, September 2015.

[23] M. O. Agyeman, Q. T. Vien, A. Ahmadinia, A. Yakovlev, K. F. Tong, and T. Mak, "A resilient 2-D waveguide communication fabric for hybrid wired-wireless NoC design," *IEEE Transactions on Parallel and Distributed Systems*, vol. 28, no. 2, pp. 359–373, 2017.

[24] I. Llatser, S. Abadal, A. M. Sugrañes, A. Cabellos-Aparicio, and E. Alarcón, "Graphene-enabled wireless networks-on-chip," in *Proceedings of the 2013 1st International Black Sea Conference on Communications and Networking, BlackSeaCom 2013*, pp. 69–73, July 2013.

[25] S. Abadal, A. Mestres, M. Iannazzo, J. Solé-Pareta, E. Alarcón, and A. Cabellos-Aparicio, "Evaluating the feasibility of wireless networks-on-chip enabled by graphene," in *Proceedings of the 11th International Conference on Principles and Practices of Programming on the Java Platform: Virtual Machines, Languages, and Tools, PPPJ 2014*, pp. 51–56, September 2014.

[26] S. Abadal, E. Alarcón, A. Cabellos-Aparicio, M. Lemme, and M. Nemirovsky, "Graphene-enabled wireless communication for massive multicore architectures," *IEEE Communications Magazine*, vol. 51, no. 11, pp. 137–143, 2013.

[27] J. M. Jornet and I. F. Akyildiz, "Channel modeling and capacity analysis for electromagnetic wireless nanonetworks in the terahertz band," *IEEE Transactions on Wireless Communications*, vol. 10, no. 10, pp. 3211–3221, 2011.

[28] D. W. Matolak, S. Kaya, and A. Kodi, "Channel modeling for wireless networks-on-chips," *IEEE Communications Magazine*, vol. 51, no. 6, pp. 180–186, 2013.

[29] D. M. Pozar, *Microwave Engineering*, Wiley, 4th edition, 2012.

[30] L. S. Rothman, I. E. Gordon, Y. Babikov et al., "The HITRAN2012 molecular spectroscopic database," *Journal of Quantitative Spectroscopy & Radiative Transfer*, vol. 130, pp. 4–50, 2013.

[31] Y. P. Zhang, Z. M. Chen, and M. Sun, "Propagation mechanisms of radio waves over intra-chip channels with integrated antennas: Frequency-domain measurements and time-domain analysis," *IEEE Transactions on Antennas and Propagation*, vol. 55, no. 10, pp. 2900–2906, 2007.

[32] T. S. Rappaport, *Wireless Communications: Principles and Practice*, Prentice Hall, 2nd edition, 2002.

[33] A. Goldsmith, *Wireless Communications*, Cambridge University Press, New York, NY, USA, 2005.

[34] R. M. Goody and Y. L. Yung, *Atmospheric Radiation: Theoretical basis*, Oxford University Press, 2nd edition, 1989.

[35] I. S. Gradshteyn and I. M. Ryzhik, *Table of Integrals, Series, and Products*, Academic Press, Cambridge, Mass, USA, 7th edition, 2007.

[36] T. G. Kyle, *Atmospheric Transmission, Emission and Scattering*, Pergamon, Oxford, New York, NY, USA, 1991.

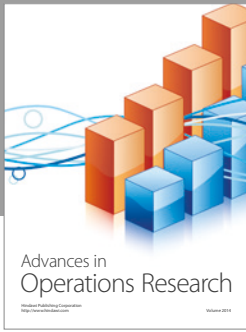 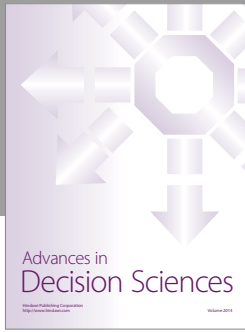 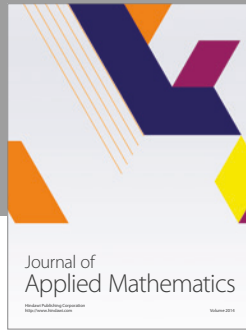 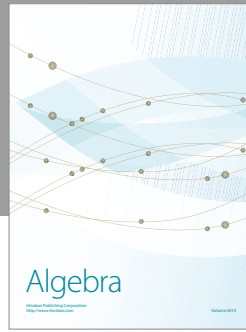 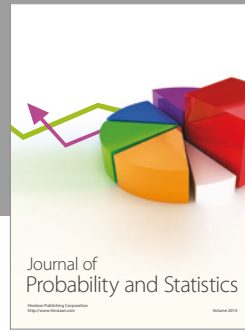
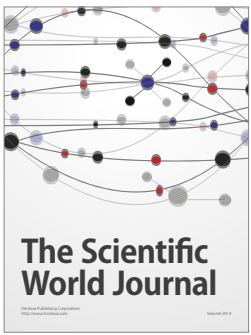 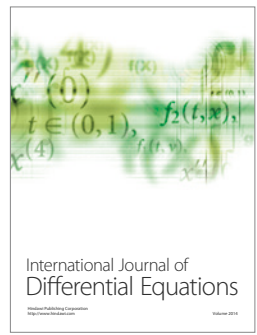
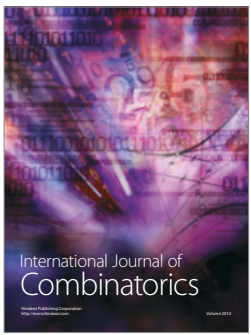 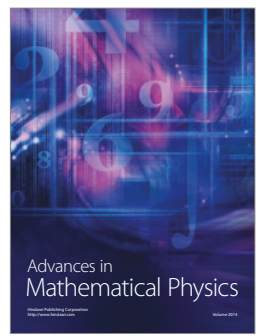
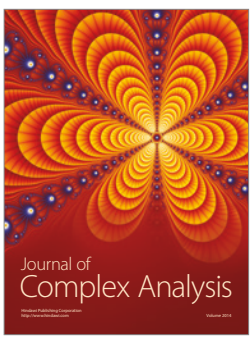 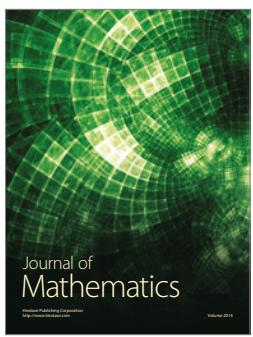 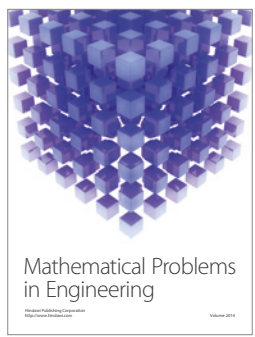 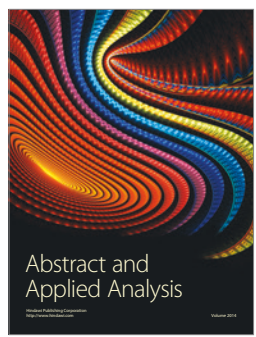 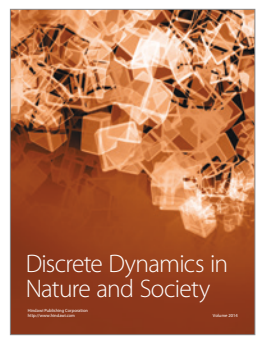
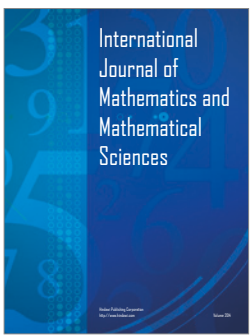 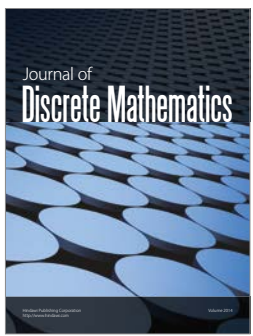 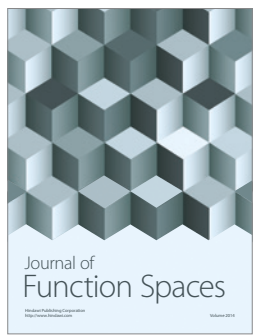 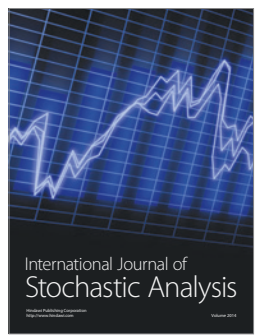 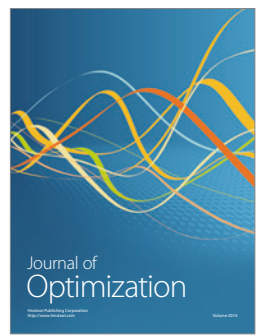

Submit your manuscripts at
https://www.hindawi.com